\documentstyle[twocolumn,graphicx,psfrag,amssymb,amsmath,eqsecnum,pra,aps]{revtex}

\def\be#1\ee{\begin{equation}#1\end{equation}}

\newcounter{theoremnr}
\newenvironment{theorem}%
   {\refstepcounter{theoremnr}%
    \begin{trivlist}%
    \item[\hskip\labelsep{\bf Theorem~\arabic{theoremnr}}]\itshape}%
   {\end{trivlist}}

\newcounter{definitionnr}
\newenvironment{definition}%
   {\refstepcounter{definitionnr}%
    \begin{trivlist}%
    \item[\hskip\labelsep{\bf Definition~\arabic{definitionnr}}]\itshape}%
   {\end{trivlist}}

\newenvironment{proof}%
   {\begin{trivlist}%
    \item[\hskip\labelsep\itshape Proof:]\itshape}%
   {\hspace*{\fill}$\square$\end{trivlist}}

\newcommand{\bea}{\begin{eqnarray}}
\newcommand{\eea}{\end{eqnarray}}

\newcommand{\bmat}{\begin{pmatrix}}
\newcommand{\emat}{\end{pmatrix}}

\newcommand{\ket}[1]{{\left|{#1}\right>}}

\newcommand{\commutator}[2]{{\left[{#1},{#2}\right]}}

\newcommand{\diag}{{\operatorname{diag}}}

\newcommand{\trace}[1]{{\operatorname{Tr}{#1}}}

\newcommand{\vek}[1]{{\boldsymbol{#1}}}
\newcommand{\mat}[1]{{\mathsf{#1}}}
\newcommand{\op}[1]{{\hat{#1}}}

\newcommand{\Ham}{{\mathcal{H}}}
\newcommand{\Hambar}{{\bar{\mathcal{H}}}}
\newcommand{\Hamtau}[1]{{\Ham_{#1}\tau_{#1}}}

\newcommand{\sigx}{{\op{\sigma}_x}}
\newcommand{\sigy}{{\op{\sigma}_y}}
\newcommand{\sigz}{{\op{\sigma}_z}}
\newcommand{\sigi}{{\op{\sigma}_i}}
\newcommand{\sigj}{{\op{\sigma}_j}}
\newcommand{\sigk}{{\op{\sigma}_k}}

\newcommand{\oa}{$O\!A$}
\newcommand{\OA}[1]{$O\!A(#1)$}

\newcommand{\durchgezogen}
     {\begin{picture}(15,10)\put(0,3){\line(1,0){15}}\end{picture}}

\newcommand{\gestrichelt}
     {\begin{picture}(21,10)\multiput(0,3)(8,0){3}{\line(1,0){5}}\end{picture}}

\newcommand{\labs}{\scriptsize}
\newcommand{\psfragWkt}
   {\psfrag{0.2}{\labs{0.2}}\psfrag{0.4}{\labs{0.4}}\psfrag{0.6}{\labs{0.6}}
    \psfrag{0.7}{\labs{0.7}}\psfrag{0.8}{\labs{0.8}}\psfrag{0.9}{\labs{0.9}}
    \psfrag{1.}{\;\labs{1}}\psfrag{0.}{\;\labs{0}}}
\newcommand{\psfrageinsbisfuenf}
   {\psfrag{0}{\labs{0}}\psfrag{1}{\labs{1}}\psfrag{2}{\labs{2}}
    \psfrag{3}{\labs{3}}\psfrag{4}{\labs{4}}\psfrag{5}{\labs{5}}}
\newcommand{\psfragWktzwei}
   {\psfrag{1.}{\;\labs{1}}\psfrag{0.992}{\,\labs{0.992}}
    \psfrag{0.994}{\,\labs{0.994}}\psfrag{0.996}{\,\labs{0.996}}
    \psfrag{0.998}{\,\labs{0.998}}}
\newcommand{\psfragzweierpotenzen}
   {\psfrag{1}{\labs{1}}\psfrag{2}{\labs{2}}\psfrag{4}{\labs{4}}
    \psfrag{8}{\labs{8}}\psfrag{16}{\labs{16}}}

\newcommand{\iu}{{\mathrm{i}}}

\begin{document}

\draft

\title{Suppression of arbitrary internal coupling in a quantum register}
\author{Marcus Stollsteimer and G\"unter Mahler}
\address{Institut f\"ur Theoretische Physik,
         Universit\"at Stuttgart, Pfaffenwaldring~57,
         70550~Stuttgart, Germany}
\date{\today}


\maketitle
\begin{abstract}
For the implementation of a quantum computer it is necessary
to exercise complete control over the Hamiltonian of the
used physical system. For NMR quantum computing the effectively acting
Hamiltonian can be manipulated via pulse sequences.
Here we examine a register consisting of $N$ selectively addressable spins
with pairwise coupling between each spin pair. We show that complete
decoupling of the spins is possible independent of the particular form
of the spin-spin interaction. The proposed method based on
orthogonal arrays is efficient in the
sense that the effort regarding time and amount of pulses increases
only polynomially with the size~$N$ of the register.
However, the effect of external control errors in terms of
inaccurate control pulses eventually limits the achievable precision.
\end{abstract}



\narrowtext
\section{Introduction}

Couplings within a quantum register may be considered a necessary evil:
the two-qubit operations needed for the realization of quantum computing
\cite{DiVincenzo:QuantumComputation}
rely on interactions between different qubits.
For fast gate operations even strong couplings are desirable.
On the other hand, residual interactions, that are always present,
disturb the system and lead to unwanted evolution, so that stored
information may get lost. This kind of ``hardware'' problems
exist even in situations where external (classical) contacts are
supposed to control pertinent parameters of the network Hamiltonian
\cite{LossDiVincenzo:QuantumDot,Kane:SiliconBasedQC}.
They are particularly relevant for conventional NMR implementations
\cite{GershenfeldChuang:NMR-QC,Cory:NMR-QC},
where couplings cannot simply be turned on and off, but are predetermined
by the respective Hamiltonian.

The interplay between local disorder and interactions has recently been
investigated with respect to the possible emergence of the so-called
``quantum chaos'' \cite{GeorgeotShepelyansky:QuantumChaosBorder}.
As expected, disorder helps to avoid the delocalization
of local register states. On the other hand, (static) disorder itself
presents a kind of control loss, which is not welcome. We will include
some aspects thereof via faulty gate operations.

Fortunately, by using sophisticated pulse sequences a wide range of control
can be exerted on the system.
Coupling terms can be removed in an NMR spin echo experiment
by applying so-called refocusing pulses, essentially
reversing the time evolution.
This method can be extended to the task of removing selected or even
all coupling terms in a quantum register of $N$ qubits.
Such a scheme has been described by Linden et al.~\cite{Linden:Refocusing}.
Unfortunately it lacks efficiency.
Efficient schemes have been proposed by
Leung et al.~\cite{Leung:Recoupling} and
Jones and Knill~\cite{JonesKnill:Refocusing}, respectively.
However, these studies focus on the special model of weak
scalar coupling only.
For other coupling models it remains open whether
there exist generalized schemes, let alone efficient ones.
In this paper we will demonstrate that, indeed, any interaction can
be suppressed efficiently, regardless of the specific coupling model, and even
without knowledge of the underlying Hamiltonian.

Perfect decoupling in general requires fast application
of many pulses. In a real implementation the pulses will never be ideal,
which will impair the decoupling.
Quantum error correction methods could be used, but they require
supplementary resources in the form of auxiliary qubits.
Here, we want to probe the limitations to the given
method that arise from faulty gate operations without considering
error correction methods.

\section{Model and Average Hamiltonian}

We consider a system of $N$ coupled spin-1/2 particles (qubits)
with the general Hamiltonian model
\be\label{general-coupling-Ham}
\Ham = \sum_{\mu=1}^N \sum_{i=1}^3 \omega_i^{(\mu)} \sigi^{(\mu)} +
       \sum_{\mu<\nu}^N \sum_{i,j=1}^3
               H_{ij}^{\mu\nu}\, \sigi^{(\mu)} \!\otimes \sigj^{(\nu)},
\ee
where the $\sigi$ ($i=1,2,3$ or $x,y,z$) denote the
Pauli operators. The terms linear in those are the Zeeman terms specified
by the 3-dimensional vectors $\vek{\omega}^{(\mu)}=(\omega_i^{(\mu)})$.
The interaction between two qubits $\mu$ and $\nu$
is described by the coupling tensor $\mat{H}^{\mu\nu}=(H_{ij}^{\mu\nu})$.
We will call couplings ``diagonal'' when they are represented by a
diagonal coupling tensor \mbox{$\mat{H}=\diag(J_x,J_y,J_z)$}.
Examples are the strong and weak scalar coupling,
with \mbox{$J_x=J_y=J_z=:J$}
and \mbox{$J_x=J_y=0$},
respectively.

We will now subject the system to a cyclic pulse train.
The pulses are assumed to be infinitely short, so that they can be
represented by instantaneous operations $\op{P}_k$.
Each pulse is followed by a free evolution period of duration $\tau_k$.
A sequence consisting of $n$ pulses will thus be denoted by
\be\label{sequence}
(\op{P}_1,\tau_1,\dots,\op{P}_n,\tau_n),
\ee
where the pulses are applied from left to right.
For the further analysis we use the formalism of
average Hamiltonian (AH) theory
\cite{Slichter:NMR,Ernst:NMR,Haeberlen:HighResNMR}.
The time evolution operator is given by
$
\op{U}(t_c)= \prod_{k=1}^n e^{-\iu\Ham\tau_k} \op{P}_k ,
$
where $t_c := \sum_{k=1}^n \tau_k$ is the cycle time
(all products are ordered with increasing $k$ from right to left).
For a cyclic sequence (defined by $\prod_{k=1}^n \op{P}_k=\op{1}$),
we can write this as
$\op{U}(t_c)=\prod_{k=1}^n
 [(\prod_{l=1}^k \op{P}_l)^{-1} e^{-\iu\Ham\tau_k} (\prod_{l=1}^k \op{P}_l) ]$.
Using the identity
$\op{A}^{-1}\exp(\op{B})\,\op{A}=\exp(\op{A}^{-1}\op{B}\,\op{A})$
we get
\be\label{time-evolution-operator}
 \op{U}(t_c)=\prod_{k=1}^n e^{-\iu\Hamtau{k}} .
\ee
The operators $\Ham_k$ are the Hamiltonians in the so-called
``toggling frame'' \cite{Ernst:NMR}
\be\label{toggling-frame-Ham}
\Ham_k := \op{Q}_k^{-1} \Ham \op{Q}_k,
\quad\text{with }
\op{Q}_k := \prod_{l=1}^k \op{P}_l .
\ee
By expanding the exponential in~(\ref{time-evolution-operator}) and
collecting terms of equal order in $\Hamtau{k}$ the time evolution operator
can be written as
\be
\op{U}(t_c)=e^{-\iu\left(\Hambar+\Hambar^{(1)}+\Hambar^{(2)}+\cdots\right)t_c},
\ee
where the operators $\Hambar$, $\Hambar^{(1)}$, $\dots$ are the
average Hamiltonians of increasing order.
In the remainder of this paper we will only use pulse sequences consisting
of equally spaced pulses ($\tau_k=\tau$).
For such a sequence with $n$ pulses (``$n\tau$-sequence'')
the leading order terms of the AH expansion are given by
\bea\label{AH}
\Hambar       &=& \frac{1}{n} \sum_{k=1}^n \Ham_k , \\
\Hambar^{(1)} &=& -\frac{\iu t_c}{2 n^2}
                  \sum_{j>k} \commutator{\Ham_j}{\Ham_k} .
    \label{AHcorr}
\eea
(For a more detailed derivation refer to
App.~B of \cite{Haeberlen:HighResNMR}.)
The time evolution can therefore be described approximately
by the average of the toggling frame Hamiltonians, $\Hambar$,
in the following simply referred to as ``the average Hamiltonian''.
This holds exactly for vanishing commutators between all $\Ham_k$.

For a given time period $k$, the pulse sequence will lead to the
transformation $\op{Q}_k=\bigotimes_{\mu=1}^N\op{Q}_k^{(\mu)}$,
where $\op{Q}_k^{(\mu)}$ is the transformation on qubit~$\mu$.
The applied pulses, and hence the operators $\op{Q}_k^{(\mu)}$,
are unitary transformations.
Any unitary transformation
in $SU(2)$ can be represented by ordinary 3-dimensional
rotation matrices $\mat{R}=(R_{ij})$ via the relation
\be\label{sigma-trafo}
\op{Q}^\dag \sigi \op{Q} = \sum_{j=1}^3 R_{ij} \sigj ,
\ee
so that any pulse sequence can be characterized by a
set of matrices $\mat{R}_k^{(\mu)}$.

The resulting average Hamiltonian can be written in the same form
as the original system Hamiltonian (\ref{general-coupling-Ham});
it is straightforward to verify from
(\ref{toggling-frame-Ham}) and (\ref{sigma-trafo})
that the vector representing the Zeeman part for qubit $\mu$ is given by
\be\label{average/transformed-Zeeman-vector}
\bar{\vek{\omega}}^{(\mu)} = \frac{1}{n} \sum_{k=1}^n \vek{\omega}_k^{(\mu)}
              = \frac{1}{n} \sum_{k=1}^n
                \bigl(\mat{R}_k^{(\mu)}\bigr)^T \vek{\omega}^{(\mu)} ,
\ee
and the $3\times 3$ tensor representing the coupling part for a
given pair $\mu$, $\nu$ by
\be\label{average/transformed-coupling-tensor}
\bar{\mat{H}}^{\mu\nu} = \frac{1}{n} \sum_{k=1}^n \mat{H}_k^{\mu\nu}
              = \frac{1}{n} \sum_{k=1}^n
                \bigl(\mat{R}_k^{(\mu)}\bigr)^T \mat{H}^{\mu\nu}\, \mat{R}_k^{(\nu)} .
\ee

Our goal is to effectively remove all interactions by applying
a proper pulse sequence. The more general task of generating
an evolution corresponding to selected couplings being turned on or off
requires only minor modifications.
Specifically, we want to achieve ``first order decoupling'', meaning
that the average Hamiltonian vanishes. For now higher order terms in the
AH expansion will be neglected.
Also, the Zeeman interactions shall be excluded for the time being.

\section{Decoupling for 2-Qubit Network}

We consider at first the case of two qubits.
Ignoring Zeeman interactions we are left with a single
interaction term represented by the tensor $\mat{H}^{12}=:\mat{H}$.
First order decoupling is then achieved by a pulse
sequence resulting in $\bar{\mat{H}}=0$.

Two principally different kinds of pulses can be used, spin-selective
and nonselective ones.
From~(\ref{average/transformed-coupling-tensor}) the following
theorem on the selectivity of the required decoupling pulses
can be derived:
\begin{theorem}
A coupling represented by a tensor $\mat{H}$
with $\trace{\mat{H}}\ne 0$ cannot be decoupled with nonselective pulses.
\end{theorem}
\begin{proof}
For nonselective pulses $(\mat{R}_k^{(1)}=\mat{R}_k^{(2)})$
the toggling frame coupling tensor is
an orthogonal transformation of the original coupling tensor,
therefore $\trace{\mat{H}_k}=\trace{\mat{H}}=\trace{\bar{\mat{H}}}$.
For decoupling one needs $\bar{\mat{H}}=0$, and hence $\trace{\mat{H}}$
must vanish. (This also holds for sequences with different $\tau_k$,
where (\ref{average/transformed-coupling-tensor})
is then given by a weighted average.)
\end{proof}

In the case of, e.\,g., direct dipolar couplings
(with \mbox{$\mat{H}\propto\diag(1,1,-2)$} in the high-field
approximation) this condition is fulfilled and nonselective
sequences like the well-known WHH-4 \cite{Waugh:HighresNMR} can be applied.
On the other hand, the widely discussed strong or weak scalar coupling
requires selective pulses.

We now give a specific first order decoupling sequence for
arbitrary couplings using spin-selective $\pi$~pulses
with respect to the coordinate axes $x,y,z$.
These pulses are given by the operators
$\op{X}=e^{-\iu\pi\op{\sigma}_x /2}=\iu\op{\sigma}_x$,
and analogously for $\op{Y}$ and $\op{Z}$.
The corresponding rotation matrices are
\bea
X&=&\mat{R}_x=\diag(+1,-1,-1) , \nonumber\\
Y&=&\mat{R}_y=\diag(-1,+1,-1) , \\
Z&=&\mat{R}_z=\diag(-1,-1,+1) , \nonumber
\eea
with the multiplication table shown in Table~\ref{mult-table}.
The decoupling is based on the following property
\be\label{trafos-sum}
a_0 I + a_1 X + a_2 Y + a_3 Z = 0
\quad\Leftrightarrow\quad
a_i = a ,
\ee
where $I$ denotes the identity matrix.
A weighted sum of the matrices $I$, $X$, $Y$, $Z$ vanishes if and
only if the weight factors are equal.

Following \cite{Leung:Recoupling} a pulse sequence will be written
as a matrix of transformations 
$I$, $X$, $Y$, and $Z$ specifying the corresponding $\op{Q}$
or $\mat{R}$ on a specific qubit for a certain time interval
of duration $\tau$.
The $n$ time intervals correspond to the columns (from left to right)
and the different qubits $\mu=1,2,\dots,N$ to different rows.
We may thus speak of an $n$-dimensional row vector $\vek{R}^{(\mu)}$,
the elements of which are the ordered sequence of transformation matrices
for given $\mu$, $\mat{R}_k^{(\mu)}$, $k=1,2,\dots,N$.
This representation gives more direct insight into what is happening
to the coupling Hamiltonian than the notation involving the actually
applied pulses.

Now consider the sequence
\be\label{general-decoupling}
\left(\begin{array}{c}
\vek{R}^{(1)} \\
\vek{R}^{(2)}
\end{array}\right)
=
\left(\begin{array}{*{4}{c}}
I&I&I&I \\
I&X&Z&Y
\end{array}\right) ,
\ee
which can be implemented by the following train of $\pi$~pulses on qubit~2
[cf.~(\ref{sequence})]
\be
(\tau,\op{X}^{(2)},\tau,\op{Y}^{(2)},\tau,\op{X}^{(2)},\tau,-\op{Y}^{(2)}) .
\ee
(The final $y$-pulse is applied in order to make the sequence cyclic.)
The resulting average Hamiltonian is given by
\be
\bar{\mat{H}} = \frac{1}{n}\, \mat{H} \,\sum_k \mat{R}_k^{(2)}
              = \frac{1}{n}\, \mat{H} \,(I+X+Y+Z) ,
\ee
which vanishes due to (\ref{trafos-sum}) for any coupling tensor $\mat{H}$.
This sequence thus effectively removes an arbitrary coupling between
two qubits. The Hamiltonian does not even have to be known, because the
same sequence works for any interaction.
Hence it follows:
\begin{theorem}
Any interaction between two spins can be decoupled to first
order using a sequence of spin-selective pulses.
\end{theorem}

For diagonal interaction models
(\ref{average/transformed-coupling-tensor}) becomes
\be\label{transformed-diagonal-coupling-tensor}
\bar{\mat{H}} = \frac{1}{n}\mat{H} \sum_{k=1}^n \mat{R}_k^{(1)} \mat{R}_k^{(2)}
              =: \frac{1}{n}\mat{H} \sum_{k=1}^n \mat{R}_k^{(12)} ,
\ee
because all matrices are diagonal (for the $\pi$~pulses under investigation).
Then it does not matter how the transformations are ``distributed'' between
the two qubits, because only the product needs to be considered.
Thus alternative sequences are possible, which will be particularly
interesting for the $N>2$ case discussed later.
The sequence
\be\label{diagonal-decoupling}
\left(\begin{array}{c}
\vek{R}^{(1)} \\
\vek{R}^{(2)}
\end{array}\right)
=
\left(\begin{array}{*{4}{c}}
I&X&X&I \\
I&I&Y&Y
\end{array}\right)
\ee
leads to the same transformations
of the coupling tensor as (\ref{general-decoupling}) and thus also
decouples the interaction.
Furthermore, it has the additional feature of removing the Zeeman terms,
for the typical case that they only consist of
a $z$-component, $\omega_z\op{\sigma}_z$.
The corresponding pulses applied both on qubit 1 and 2 are given by
\be\label{diagonal-decoupling-pulses}
(\tau,\op{X}^{(1)},\tau,\op{Y}^{(2)},\tau,\op{X}^{(1)},\tau,\op{Y}^{(2)}) .
\ee

The sequential order of the transformations clearly has no influence on
$\bar{\mat{H}}$ and thus on the first order decoupling.
For the above mentioned sequences it is chosen so that no $z$-pulses
are needed, which usually cannot be implemented in conventional NMR.
The higher order terms, however, are affected by the order.

\section{Decoupling for $N$-Qubit Network}

We now turn to a system consisting of $N$ qubits.
It often suffices to take only nearest and, possibly,
next nearest neighbors into account. We will come back to this issue later.
For now we discuss the most general case of a completely coupled
system, i.\,e.\ each qubit is coupled to all the other ones.
From (\ref{average/transformed-coupling-tensor}) it is evident that
each pair interaction $\mat{H}^{\mu\nu}$ can be treated separately,
essentially leading back to the $N=2$ case.
So, we can use the sequence for $N=2$ as a starting point.
An $n\tau$-decoupling sequence for $N$ qubits will be represented by an
$N\times n$ matrix. Decoupling means elimination of all $\mat{H}^{\mu\nu}$.
The task is to find a matrix that yields a valid $N=2$ decoupling sequence
for any pairing of rows.

\subsection{Simple Scheme}

The simplest approach is to recursively nest the sequence
(\ref{general-decoupling})
in analogy to the method in \cite{Linden:Refocusing}.
When a qubit is added to a valid sequence, it will be exposed
to a complete unit cycle during each time interval of the
initial sequence.
For 3~qubits we thus get:
\be
\setlength{\arraycolsep}{1.7pt}
\left(\begin{array}{*{16}{c}}
I&I&I&I&I&I&I&I&I&I&I&I&I&I&I&I \\
I&I&I&I&X&X&X&X&Z&Z&Z&Z&Y&Y&Y&Y \\
I&X&Z&Y&Y&Z&X&I&I&X&Z&Y&Y&Z&X&I
\end{array}\right) .
\ee
While the additional qubit will undergo one $4\tau$ unit cycle,
the transformations on the other qubits are constant,
and it will therefore decouple.
However, this scheme is not efficient in terms of the time intervals $\tau$
and pulses needed, which both scale as $4^{N-1}$
because the sequence has to be prolonged by a factor~4
for each added qubit.
The advantage is that at each time interval only one pulse has to be applied.
This is why every second unit cycle is reversed.
Furthermore, the almost realized mirror symmetry ($\Ham_k=\Ham_{n+1-k}$)
is expected to reduce the higher order correction terms:
for a symmetric sequence, all correction terms of odd order
vanish \cite{Haeberlen:HighResNMR}.

\subsection{Efficient Scheme Using Orthogonal Arrays}
\label{sec:eff-scheme}

Much more efficient schemes can be constructed using the notion
of orthogonal arrays (\oa), which have been extensively studied in the
context of combinatorics, design theory, and error-correcting codes.
A comprehensive treatment can be found in
\cite{Hedayat:OrthogonalArrays}.
First we want to give a definition of orthogonal arrays and then
demonstrate how they are related to the decoupling problem.

Let $S$ be a set of $s$ symbols.
Then an orthogonal array is defined by
\begin{definition}\label{def:OA}
A $k\times \lambda s^t$ array $A$ with entries from $S$ is said to be an
orthogonal array with $s$ levels, strength $t$, and index $\lambda$
$(0\le t\le k)$ if every $t\times \lambda s^t$ subarray of $A$
contains each $t$-tuple based on $S$ exactly $\lambda$ times as a column.
It will be denoted by \OA{\lambda s^t, k, s, t}. 
\end{definition}

For our purposes we will only regard the case $t=2$ and $s=4$.
Usually the symbols are chosen as the numbers 0, 1, 2, 3.
We will interpret them as the $\pi$ rotation matrices
including the identity matrix,
so that $S=\{I,X,Y,Z\}$.
As an example regard the following \OA{16, 5, 4, 2}
\be\label{OA-16-5}
\setlength{\arraycolsep}{1.7pt}
\left(\begin{array}{*{16}{c}}
I&I&I&I&X&X&X&X&Y&Y&Y&Y&Z&Z&Z&Z \\
I&X&Y&Z&I&X&Y&Z&I&X&Y&Z&I&X&Y&Z \\
I&X&Y&Z&X&I&Z&Y&Y&Z&I&X&Z&Y&X&I \\
I&X&Y&Z&Y&Z&I&X&Z&Y&X&I&X&I&Z&Y \\
I&X&Y&Z&Z&Y&X&I&X&I&Z&Y&Y&Z&I&X
\end{array}\right) .
\ee
If you pick any two rows (e.\,g.\ row 2 and 5), each
of the $s^t=16$ possible 2-dimensional column vectors appears
exactly once, hence $\lambda=1$.
Definition~\ref{def:OA} implies further that every row contains
each symbol exactly $\lambda s=4$ times.
As it turns out, these properties are sufficient for
the array to be a valid decoupling sequence:
\begin{theorem}
Any \OA{16\lambda, k, 4, 2} can be used to decouple $k+1$ qubits
within $n=16\lambda$ time intervals.
\end{theorem}
\begin{proof}
Pick two rows $\mu$ and $\nu$ of the \oa\ and consider the corresponding
average Hamiltonian
according to (\ref{average/transformed-coupling-tensor}),
$\bar{\mat{H}}^{\mu\nu} = \frac{1}{n} \sum_{j=1}^n
 \mat{R}_j^{(\mu)}\, \mat{H}^{\mu\nu}\, \mat{R}_j^{(\nu)}$,
$n=16\lambda$.
Each of the $s^t=16$ combinations
$(\mat{R}_j^{(\mu)}$,$\mat{R}_j^{(\nu)})=(I,I),(I,X),\dots,(Z,Z)$
occurs exactly $\lambda$ times.
We thus get
$
\bar{\mat{H}}^{\mu\nu}=\frac{1}{16}
                    (I+X+Y+Z)\,\mat{H}^{\mu\nu}\,(I+X+Y+Z)
$,
which vanishes for any $\mat{H}^{\mu\nu}$.
We can further add an additional $(k+1)$th row to the array, consisting
only of $I$'s. As every row of the \oa\ contains
an equal number $\lambda s$ of each symbol, the AH involving the
$(k+1)$th qubit is $\bar{\mat{H}}^{\mu,k+1}=\frac{1}{4}
                    \mat{H}^{\mu,k+1}\, (I+X+Y+Z)=0$.
Thus, the \oa\ can be used to decouple $k+1$ qubits.
\end{proof}

Using large enough \oa s ($k+1\ge N$), decoupling sequences for an
arbitrary number $N$ of qubits can thus be constructed.
We note that simultaneous pulses on different qubits
are necessary for this scheme to work.
The essential question is for which sets of parameters
$\lambda, k, s, t$ an \oa\ exists at all.
Specifically, we want to find arrays with $k$ as large as possible,
in order to decouple many qubits with the least effort.

Bose and Bush \cite{BoseBush:OrthogonalArrays} have given a recursive
construction scheme for arrays of strength 2,
providing \oa s with large values $k$.
From their Theorem~4 follows the existence of all
\OA{n, k, 4, 2} with
\be\label{BoseBush}
n=2^{u+4},\quad
k=\left\{
\begin{array}{ll}
\frac{1}{3}(n-1), & u=0,2,4,\dots\\[2pt]
\frac{1}{3}(n-5), & u=1,3,5,\dots
\end{array}
\right. .
\ee
An example of the underlying construction scheme will be given in the
appendix.
This guarantees the existence of the arrays
\OA{16,5,4,2}, \OA{32,9,4,2}, \OA{64,21,4,2}, \OA{128,41,4,2}, and so on.
[Note that this construction does not yield all possible \oa s,
there exists, for example, an \OA{48,13,4,2}.]
From (\ref{BoseBush}) we can derive a lower bound on
the efficiency of the scheme:
An $(n=2^{u+4})\tau$-sequence can be used to decouple
up to $N=\frac{1}{3}(n\pm 2)$ qubits (the sign depending on $u$ being
even or odd).
It follows that $N$ qubits ($N \ge 3$) can be decoupled by a sequence
consisting of $n=cN-2$ time intervals, where $3\le c \le 6$.
The number of pulses that are needed for the sequence is less than $N(cN-2)$.
Hence it is possible to decouple (to first order) any interaction
with linearly increasing effort in time:
the scheme is efficient.

\subsection{Efficient Scheme for Diagonal Couplings}

For diagonal couplings there exists an even more efficient scheme.
When the transformations on two qubits $\mu$ and $\nu$ are given by the
row vectors $\vek{R}^{(\mu)}$ and $\vek{R}^{(\nu)}$
[cf.~(\ref{general-decoupling})], the average Hamiltonian is,
in analogy to (\ref{transformed-diagonal-coupling-tensor}),
\be
\bar{\mat{H}}^{\mu\nu}
=\frac{1}{n} \mat{H}^{\mu\nu}
 \sum_{k=1}^n \mat{R}_k^{(\mu)}\mat{R}_k^{(\nu)}
=:\frac{1}{n} \mat{H}^{\mu\nu}
 \left(\vek{R}^{(\mu)}\!\cdot\vek{R}^{(\nu)}\right) .
\ee
It vanishes whenever the ``scalar'' product
$\vek{R}^{(\mu)}\!\cdot\vek{R}^{(\nu)}$ (which is really tensor valued)
equals zero.
This leads with (\ref{trafos-sum}) to the
following decoupling criterion:
\begin{theorem}\label{theorem-scalar-prod}
A pulse sequence represented by a matrix of $N$ rows
can be used to decouple $N$ qubits with diagonal coupling
between any qubit pair if the scalar products between different
row vectors vanish. This is equivalent to the condition that the
vector of the element-wise products of two rows contains each of the
elements $I, X, Y, Z$ equally often.
\end{theorem}

The sequences already discussed,
like e.\,g.\ (\ref{OA-16-5}),
fulfill this condition, but there
exist suitable matrices with a smaller number of columns, leading to
even shorter sequences.
These matrices are closely related to the so-called
difference schemes or matrices
\cite{Hedayat:OrthogonalArrays}, which play an important role in the
construction of the \oa s discussed in the previous section.
Difference schemes are defined over a finite Abelian group
of $s$ elements with a binary operation $+$, $(\mathcal{A},+)$.

\begin{definition}\label{def:D}
An $r\times c$ array $D(r,c,s)$ with entries from $\mathcal{A}$ is
called a difference scheme based on $(\mathcal{A},+)$,
if for all $\mu$ and $\nu$ with $1\le \mu,\nu\le r$,
$\mu\ne\nu$, the vector difference between the $\mu$th and $\nu$th rows
contains every element of $\mathcal{A}$ equally often.
\end{definition}

Relevant here are difference schemes over the
additive group of the Galois field $G\!F(s)$
(cf.\ App.~A of \cite{Hedayat:OrthogonalArrays})
with $s=4$ elements:
\begin{theorem}
A $D(r,c,4)$ over $(G\!F(4),+)$ is equivalent to
a $c\tau$-decoupling sequence for $N=r$ qubits.
\end{theorem}
\begin{proof}
The group of $\pi$ rotations, $(\{I,X,Y,Z\},*)$, is isomorphic
to $(G\!F(4),+)$, which is evident from Table~\ref{mult-table}.
So the elements $I$, $X$, $Y$, $Z$ can be identified with $0$, $1$, $2$, $3$,
and the matrix product with the operation $+$.
Each element is its own inverse, so that the difference
of two rows equals the sum.
Then the defining property for difference schemes is equivalent to the
condition of Theorem~\ref{theorem-scalar-prod}.
\end{proof}

For difference schemes one has \mbox{$r\le c$}
\cite{Jungnickel:DifferenceMatrices}, so that at most \mbox{$N=n$} qubits
can be decoupled with an $n\tau$-sequence.
An $r\times r$ array $D(r,r,s)$ is also called generalized Hadamard matrix.

Multiplying a column with a specific element leaves the scalar products
invariant, so that the matrices can be brought into a normalized form,
with only $I$'s in the first row.
The other rows then have to consist of permutations of equal numbers
of $I$, $X$, $Y$, $Z$, so that from the outset only sequences with
$n=4u$ ($u\in\mathbb{N}$) time intervals are applicable.
By trying all combinations we have found the following matrices for
decoupling of up to 4 or 8 qubits, respectively.
\be
M_4 =
\left(\begin{array}{*{4}{c}}
I&I&I&I \\
I&X&Y&Z \\
I&Y&Z&X \\
I&Z&X&Y
\end{array}\right) ,
\ee
\be
M_8 =
\left(\begin{array}{*{8}{c}}
I&I&I&I&I&I&I&I \\
I&I&X&X&Y&Y&Z&Z \\
I&X&Y&Z&I&X&Y&Z \\
I&X&Z&Y&Y&Z&X&I \\
I&Y&I&Y&Z&X&Z&X \\
I&Y&X&Z&X&Z&I&Y \\
I&Z&Y&X&Z&I&X&Y \\
I&Z&Z&I&X&Y&Y&X
\end{array}\right) .
\ee
We will denote such $n\times n$ decoupling matrices
corresponding to a $D(n,n,4)$ by $M_n$.
We do not show here the also existing $M_{12}$.
One is led to suspect that there may exist matrices $M_n$ for
all $n=4u$, $u\in\mathbb{N}$, which, however, is still an open question.
But at least for all $n=2^u$, $u\ge 2$, difference schemes can be
constructed using the properties of Galois fields
\cite{BoseBush:OrthogonalArrays}.
(For a more detailed discussion on the existence of difference schemes
see chap.~6 of \cite{Hedayat:OrthogonalArrays}.)
A simpler way to find matrices $M_n$ can be based on the following theorem:
\begin{theorem}
Given two decoupling matrices $M_{n_1}$, $M_{n_2}$,
their direct product $M_{n_1 n_2}=M_{n_1}\otimes M_{n_2}$ is again a
decoupling matrix.
\end{theorem}
\begin{proof}
The $(n_2(i-1)+k)$th row $(i=1,\dots,n_1; k=1,\dots,n_2)$ of the
matrix $M_{n_1 n_2}$, $(i,k)$,
derives from the $i$th row vector of $M_{n_1}$, $\vek{i}$,
and the $k$th row vector of $M_{n_2}$, $\vek{k}$, as
$
(i,k)=
\bmat  i_1 k_1 & \ldots & i_1 k_{n_2} & \ldots
     & i_{n_1} k_1 & \ldots & i_{n_1} k_{n_2} \emat
$.
The scalar product between two rows $(i,k)$ and $(j,l)$ is
$
(i,k)\cdot(j,l)
  =   i_1 j_1 (\vek{k}\cdot\vek{l}) + \dots
    + i_{n_1} j_{n_1} (\vek{k}\cdot\vek{l})
  = (\vek{i}\cdot\vek{j}) (\vek{k}\cdot\vek{l})
$.
Because all rows in $M_{n_1}$ and $M_{n_2}$ are orthogonal to each other,
$(i,k)\cdot(j,l)$ can only be different from zero for $j=i$, $l=k$.
Therefore all rows of $M_{n_1 n_2}$ are orthogonal.
\end{proof}

Now using the matrices $M_4$ and $M_8$ given above one can construct
decoupling matrices $M_{2^u}$ for any~\mbox{$u\ge 2$}.
Consequently, for $N$ qubits there exists a $(cN)\tau$-decoupling
sequence with at most $c N^2$ pulses, where $1\le c<2$.
So, for diagonal couplings the efficiency is further improved
compared to the general case of the last section.

\subsection{Zeeman terms and partially coupled systems}

Removing all Zeeman terms requires only a small modification.
From (\ref{average/transformed-Zeeman-vector}) it follows that
the Zeeman terms of a specific qubit $\mu$ are removed
if the transformations on that qubit
sum up to zero, $\sum_{k=1}^n \mat{R}_k^{(\mu)} = 0$.
This corresponds to a vanishing row sum in the matrix of
transformations.
For the given sequences this is fulfilled for all rows except
the first. Thus only the Zeeman interaction of qubit~1 is
not removed.
So, all Zeeman terms can be suppressed by using a matrix for
$N+1$ qubits and omitting the first row.

Similarly, leaving selected couplings intact rather than removing all
interaction terms takes only small alterations.
To achieve, for example, a time evolution corresponding to one
single coupling between qubits~1 and $\mu$, it is merely necessary to
use the same transformations on qubit~$\mu$ as are used on qubit~1.

In reality the couplings are usually well localized and it is sufficient
to take only close neighbors into account.
Then much simpler sequences can be used. The situation essentially
can be regarded as a coloring problem \cite{JonesKnill:Refocusing}.
For nearest-neighbor couplings the problem reduces to the $N=2$ case,
so that (\ref{general-decoupling}) or (\ref{diagonal-decoupling})
can be used, subjecting all qubits with odd number to the same
transformations and likewise for the even qubits.

\section{Higher Order Terms}

Up to now we have only considered first order decoupling, neglecting
the higher order terms in the AH expansion.
These terms can be suppressed further by repeating a given sequence with
correspondingly shorter time intervals.

Let us regard the sequence we get from repeating $m$ times a
given $n\tau$ base sequence during a given cycle time $t_c$.
It then consists of $m n$ time intervals of duration $t_c/(m n)$ each.
The average Hamiltonian is the same as for the base sequence and vanishes,
while for the first correction term we get from (\ref{AHcorr})
\be
\Hambar^{(1)} = -\frac{\iu t_c}{2 m^2 n^2}
                   \sum_{j>k}^{m n} \commutator{\Ham_j}{\Ham_k} .
\label{KorrekturTermfuermnFolge}
\ee
By using the new indices $s=1,\dots, m$ for the different subcycles and
$k=1,\dots, n$ for the time interval within a given subcycle the sum can be
written as
$
\sum_{j>k}^{m n} \commutator{\Ham_j}{\Ham_k}
= \sum_{s=s'}^m \sum_{j>k}^n \commutator{\Ham_j}{\Ham_k}+
  \sum_{s>s'}^m \sum_{j,k}^n \commutator{\Ham_j}{\Ham_k}
= m \sum_{j>k}^n \commutator{\Ham_{j}}{\Ham_{k}}
$,
implying
\be
\Hambar^{(1)} = -\frac{\iu t_c}{2 m n^2}
                   \sum_{j>k}^n \commutator{\Ham_j}{\Ham_k} .
\ee
The first correction term is thus reduced by the factor $1/m$
if the base sequence is repeated $m$ times.
In principle, by using arbitrarily fast pulses, perfect decoupling
could be realized.

\section{Example: Spin Chain}

\begin{figure}
 \begin{center}
   \psfrag{F}{$F$}\psfrag{t}{$Jt_c$}
   \psfrag{n1}{\footnotesize$m=1$}\psfrag{n4}{\footnotesize$m=4$}
   \psfragWkt
   \psfrageinsbisfuenf
   \includegraphics[width=\linewidth]{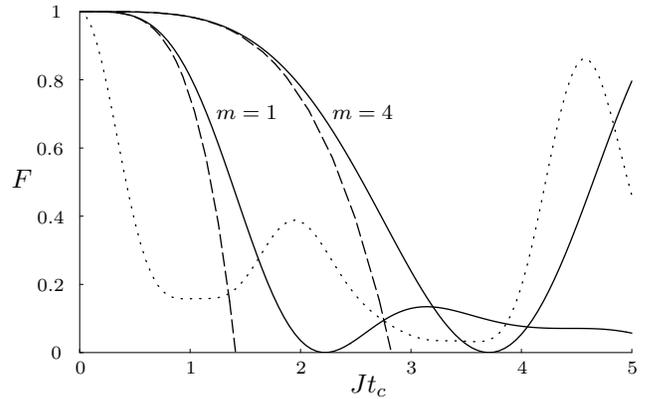}
 \end{center} 
 \caption[]{Decoupling for $N=4$ qubits,
          $\omega=0$.
          Shown is the fidelity $F(t_c)$ for
          free evolution ($\cdot\!\cdot\!\cdot\!\cdot\!\cdot$),
          decoupling with $m=1$ (\durchgezogen)
          and the corresponding approximation
          \mbox{$F=1-(Jt)^4/4$} (\gestrichelt),
          decoupling with $m=4$ (\durchgezogen) and the
          approximation \mbox{$F=1-(Jt)^4/64$} (\gestrichelt).
          }
 \label{Abb:Refoc4Spins}
\end{figure}

We will now study a simple example, a spin chain consisting
of $N$ spins with a diagonal coupling between nearest neighbors.
The corresponding Hamiltonian is [cf.~(\ref{general-coupling-Ham})]
\bea
\Ham&=& - \sum_{\mu=1}^N \omega_z^{(\mu)} \sigz^{(\mu)} \nonumber
        + \sum_{\mu=1}^{N-1}
                \left( J_x\, \sigx^{(\mu)}\!\otimes\sigx^{(\mu+1)} \right.\\
    & & \left.\mbox{}+ J_y\, \sigy^{(\mu)}\!\otimes\sigy^{(\mu+1)}
                     + J_z\, \sigz^{(\mu)}\!\otimes\sigz^{(\mu+1)} \right) .
\eea
For this interaction model,
the $N=2$ sequence for diagonal couplings
(\ref{diagonal-decoupling},\ref{diagonal-decoupling-pulses})
can be used (thus $n=4$), applying the 
$x$-pulses on all qubits with odd number and the $y$-pulses on all
qubits with even number.
The leading correction term $\Hambar^{(1)}$ does not vanish, in general:
for the respective sum of commutators in (\ref{AHcorr}) one gets by induction
\bea
\sum_{j>k}^4 \commutator{\Ham_j}{\Ham_k}&=& 8 \iu \;
 \Bigg\{
  J_x J_y \sum_{\mu=1}^{N-2}
  \left(\op{\sigma}^{(\mu)}_{xzy}+\op{\sigma}^{(\mu)}_{yzx}\right)
   \nonumber\\
 && \hspace{0mm} + \sum_{\mu=1}^{N-1}
  \left(\delta_{\mu u} a_\mu \op{\sigma}^{(\mu)}_{yx}
       +\delta_{\mu g} b_\mu \op{\sigma}^{(\mu)}_{xy} \right)
 \Bigg\} ,
\eea
with
\begin{alignat}{2}
  \op{\sigma}^{(\mu)}_{ij} &=\sigi^{(\mu)}\!\otimes\sigj^{(\mu+1)} ,
& \op{\sigma}^{(\mu)}_{ijk}
            &=\sigi^{(\mu)}\!\otimes\sigj^{(\mu+1)}\!\otimes\sigk^{(\mu+2)} ,
  \nonumber \\
  a_\mu &= \omega_z^{(\mu)} J_x + \omega_z^{(\mu+1)} J_y ,
& b_\mu &= \omega_z^{(\mu)} J_y + \omega_z^{(\mu+1)} J_x . \nonumber
\end{alignat}
As expected, the correction term vanishes for weak coupling,
where only $\sigz$ terms are involved (Ising model, $J_x=J_y=0$)
and the time evolution is given exactly by the
average Hamiltonian~(\ref{AH}).
We now define the fidelity of a given initial state $\ket{\psi}$
after the time $t_c$
\be
F(t_c) = \Big|\big<\psi\big|\op{U}(t_c)\big|\psi\big>\Big|^2 .
\ee
As an example we restrict ourselves to strong scalar
coupling (Heisenberg model, $J_x=J_y=J_z=J$) and equal chemical shifts
($\omega_z^{(\mu)}=\omega$);
taking as the initial state the state with qubit~1 excited and all
other qubits in the ground state, $\ket{\psi}=\ket{100\dots0}$,
the fidelity for small $t_c$ is approximately
\be\label{spin-chain-fidelity}
F(t_c)\approx 
\left\{\begin{array}{ll}
1-\frac{J^2 \omega^2 t_c^4}{4m^2}                 , & N=2 , \\[3pt]
1-\frac{J^2 t_c^4}{4m^2}\left[(J^2+\omega^2)(N-1)-2J^2\right], & N\ge 3 .
\end{array}\right.
\ee
In Fig.~\ref{Abb:Refoc4Spins} the undisturbed
evolution is plotted for $N=4$ qubits together with the curves for two
decoupling sequences with different pulse frequency $m$.
It can be seen how increasing $m$ improves the fidelity,
i.\,e.\ the suppression of the higher order terms.

\section{Decoupling with faulty gates}

Up to now we have only considered sequences consisting of perfect pulses,
implementing exactly a $\pi$ rotation.
This assumption is, of course, not tenable in any real physical system.
Even for ideal pulses the decoupling is generally not perfect
due to higher order terms, but as stated above, these terms can be suppressed
by increasing the pulse frequency $m$.
Yet with faulty gates this strategy is not necessarily successful,
because more errors are then introduced into the system from the outside.

In order to see what happens we study a very simple model,
two spins that interact via scalar coupling. We again use
sequence (\ref{diagonal-decoupling-pulses}) but now we assume
faulty $x$-pulses, where the rotation angle differs from $\pi$
by a small deviation $\delta$, so that
\be
\op{X}^{(1)}(\delta)=\iu\,\sigma_x^{(1)}\cos\frac{\delta}{2}
                       -\op{1}\sin\frac{\delta}{2} ,
\ee
where $\delta$ is randomly distributed with the probability distribution
$w(\delta,\sigma)=(\sqrt{2\pi}\,\sigma)^{-1} e^{-\delta^2/ \,2\sigma^2}$.
The standard deviation $\sigma$ measures the scatter of the pulses.

As a first case we consider weak coupling, with
\be
\Ham=-\omega_z^{(1)}\sigz^{(1)}-\omega_z^{(2)}\sigz^{(2)}
     +J\sigz^{(1)}\sigz^{(2)} .
\ee
Of course, in this case one would not use the above strategy because
the higher order terms vanish anyway, and increasing the number
of applied pulses will, with certainty, lead to a degraded performance.
By induction one can show analytically that after application
of $m$ decoupling cycles the fidelity for one of the product basis states
($\ket{00}$, $\ket{01}$, $\ket{10}$, $\ket{11}$)
as initial state is given by
\be\label{F-faulty-decoupling-weak}
F_\sigma(m)=\frac{1}{2}\left(1+e^{-m\sigma^2}\right) .
\ee
For increasing $\sigma$, i.\,e.\ faultier gates, the fidelity
decreases as expected.

\begin{figure}
 \begin{center}
  \psfrag{F}{\!\!\!$F_\sigma$}\psfrag{n}{$m$}
  \psfragWktzwei
  \psfragzweierpotenzen
  \includegraphics[width=\linewidth]{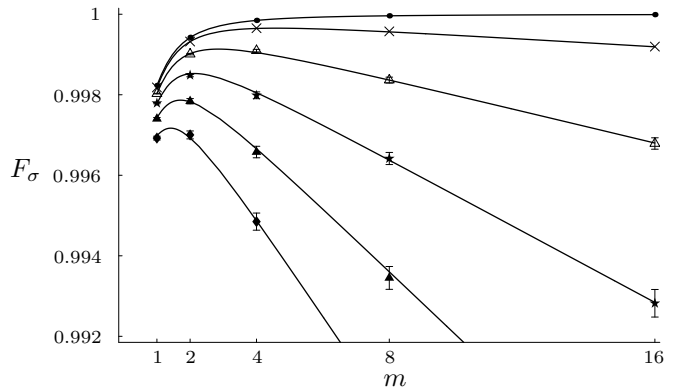}
 \end{center} 
 \caption[]
         {Decoupling for $N=2$ qubits with faulty gates,
          strong $J$-coupling.
          Shown is the fidelity $F_{\sigma}$ at $t_c=1/J$, $J=10\omega$.
          The curves correspond to $\sigma=0$~($\bullet$), 0.01~($\times$),
          0.02~($\triangle$), 0.03~($\bigstar$),
          0.04~($\blacktriangle$), 0.05~($\blacklozenge$). 
          The approximation function is
          \mbox{$F_\sigma(m)=\frac{1}{2}(1+e^{-m\sigma^2})
                                  (1-\frac{1}{400\,m^2}+\frac{1}{1369\,m^4})$};
          with correlation coefficient $0.999996$ for $\sigma=0$.}
 \label{Abb:Ungenaueentkopplungstark}
\end{figure}

For the case of strong coupling, 
\be
\Ham=-\omega\left(\sigz^{(1)}+\sigz^{(2)}\right)
      +J\,\vek{\sigma}^{(1)}\cdot\vek{\sigma}^{(2)} ,
\ee
we have resorted to a numerical calculation of the fidelity
after the time $t_c$ for the initial state $\ket{01}$, for different $\sigma$
and numbers of repetitions $m$ of base cycles.
For each set of parameters we calculated the mean over 1000 realizations
with randomly picked deviations $\delta$.
Our results are shown in Fig.~\ref{Abb:Ungenaueentkopplungstark}.
It turns out that the resulting behavior can be approximated
pretty well by the product
of the function for $\sigma=0$, approximated by (\ref{spin-chain-fidelity}),
with the ``damping'' resulting for
weak interaction (\ref{F-faulty-decoupling-weak}), yielding
\be
F_\sigma(m)\approx\frac{1}{2}\left(1+e^{-m\sigma^2}\right)
                          \left(1-\frac{1}{400\,m^2}\right) .
\ee
As one can see, we have indeed two competing effects:
for small $m$ and $\sigma$ the suppression of the
higher order corrections is dominant, whereas for larger $m$ and $\sigma$
the errors introduced by the faulty gates dominate.
In particular, there exists for a given $\sigma$ an ideal number of
iterations $m$ which yields the maximum fidelity.
The corresponding $F_{\text{max}}$ decreases thereby for increasing error.
This means that for faulty gates it is not possible anymore to
achieve perfect decoupling.
This conclusion gets increasingly significant for larger $N$:
then $F_{\text{max}}$ is further reduced by the effect
of the additional faulty gates and the rising influence of higher
order terms.

\section{Summary and Conclusions}

We have shown that it is possible to generalize refocusing methods
to systems of spins coupled by arbitrary two-spin interactions.
Thereto spin-selective operations are necessary. We have given a concrete
sequence for first order decoupling for $N=2$, and schemes for
constructing sequences for $N>2$. These schemes are efficient
regarding the needed time, which scales linearly
with the size of the system, and the amount of pulses, which scales
at most quadratically. 

However, in the general case there are higher order terms that must be
taken into account. These can be suppressed by increasing the
pulse frequency, assuming ideal gate operations.
Perfect decoupling for long times could only be achieved by infinitely fast
decoupling pulses.

These results hold for the most demanding case of an
arbitrary coupling model.
Yet for special interaction types, much more efficient schemes
are possible.
Direct dipolar couplings, for example, can be decoupled using nonselective
pulse sequences, that work independent of the system size.
For weak scalar coupling all higher order correction terms vanish,
so that the decoupling is perfect.
The sequences can also be considerably simplified for
only partly coupled systems,
like a spin chain with nearest neighbor couplings only.

In a real physical system pulses never are ideal.
For faulty gate operations perfect decoupling is no longer possible,
because of the inaccuracies introduced into the system
by the procedure itself. 

The present results are not restricted to the special task of
removing all interactions in a spin system.
Minor modifications allow specific coupling terms to be turned on or off.
Combining on- and off-periods one can even rescale the various
parameters within the original Hamiltonian.
Generally, a wide range of effective Hamiltonians can thus be invoked.
In this sense, complete decoupling is a special case of a quantum simulation
\cite{Lloyd:QuantumSimulator},
where a system is modeled that ``does nothing'',
i.\,e.\ any initial state remains unchanged.
Such a system is, in general, quite challenging to realize
and may thus be considered a serious testing scenario
for quantum dynamical control.

\section*{Acknowledgements}

We thank
J.~Gemmer,
I.~Kim,
A.~Otte,
F.~Tonner,
and
T.~Wahl
for fruitful discussions.

\appendix

\section*{Recursive Construction of Orthogonal Arrays}
\label{app:oa}

This construction is due to Bose and Bush
\cite{BoseBush:OrthogonalArrays}.
Examples can also be found in \cite{Hedayat:OrthogonalArrays}.
\begin{enumerate}
\item
Construct an \OA{16\lambda,4\lambda,4,2},
$\lambda=2^u$, $u\in\mathbb{N}$, as follows:
Take an $M_{4\lambda}$ [a $D(4\lambda,4\lambda,4)$];
then the \oa\ $A_0$ is given by
\[
A_0 = M_{4\lambda}\otimes
\left(\begin{array}{*{4}{c}}
  I & X & Y & Z \\
\end{array}\right) .
\]
\item
If possible, construct an \oa\ $A_1$ with
$\lambda_1=\lambda/4$, according to 1.
Repeat each column 4 times, yielding
\[
A_1'=A_1\otimes
\left(\begin{array}{*{4}{c}}
  I & I & I & I \\
\end{array}\right) .
\]
If possible, repeat this step with an \oa\ $A_2$ where $\lambda_2=\lambda/4^2$,
and form an $A_2'$ by repeating each column $4^2$ times, and so on.
Append the matrices $A_1'$, $A_2'$, $\ldots$ to $A_0$.
\item
Add a row consisting of $4\lambda$ $I$'s, $4\lambda$ $X$'s, and so on.
\end{enumerate}
We now have an \OA{16\lambda,k,4,2} with $\lambda=2^u$, $u\in\mathbb{N}$,
and $k=4\lambda+4\lambda_1+\cdots+1$ rows,
or
\[
k=\frac{\lambda(4^{c+1}-1)}{4^c-4^{c-1}}+1,
\quad
c=\left\{
\begin{array}{ll}
u/2, & u=0,2,\dots\\
(u-1)/2, & u=1,3,\dots
\end{array}
\right. .
\]

The \oa~(\ref{OA-16-5}) has been constructed in this way,
using $M_4$. A $\lambda_1=1/4$ does not exist,
so step~2 was omitted, resulting in $k=4+1=5$.
An \OA{32,9,4,2} can be constructed using $M_8$.
For an \OA{64,21,4,2} we construct $A_0$ using $M_{16}=M_4\otimes M_4$ and
$A_1$ using $M_4$, so $k=16+4+1=21$.


\pagebreak

\begin{table}
\caption{Multiplication table for the matrices $I, X, Y, Z$
and addition table for $G\!F(4)$}
\label{mult-table}
\[
\begin{array}{c|cccc}
*&I&X&Y&Z \\ \hline
\rule{0pt}{10pt}
I&I&X&Y&Z \\
X&X&I&Z&Y \\
Y&Y&Z&I&X \\
Z&Z&Y&X&I
\end{array}
\qquad
\begin{array}{@{\hspace{3pt}}c@{\hspace{4pt}}|@{\hspace{5pt}}*{3}{c@{\hspace{9pt}}}c@{\hspace{5pt}}}
+&0&1&2&3 \\ \hline
\rule{0pt}{10pt}
0&0&1&2&3 \\
1&1&0&3&2 \\
2&2&3&0&1 \\
3&3&2&1&0
\end{array}
\]

\end{table}

\end{document}